# Only T₃-AI can reach human-level intelligence: A variety argument

Danko Nikolić

- Department of Neurophysiology, Max Planck Institute for Brain Research, Deutschordenstraße 46, D-60528 Frankfurt/M, Germany
- Frankfurt Institute for Advanced Studies (FIAS), Ruth-Moufang-Straße 1, D-60438 Frankfurt/M, Germany
- Ernst Strüngmann Institute (ESI) for Neuroscience in Cooperation with Max Planck Society, Deutschordenstraße 46, D-60528 Frankfurt/M, Germany
- Department of Psychology, Faculty of Humanities and Social Sciences, University of Zagreb, Croatia

Correspondence:
Danko Nikolić
Max-Planck Institute for Brain Research
Deutschordenstr. 46
60528 Frankfurt am Main
email: danko.nikolic@gmail.com

**Abstract**

The recently introduced theory of practopoiesis offers an account on how adaptive intelligent systems are organized. According to that theory biological agents adapt at three levels of organization and this structure applies also to our brains. This is referred to as tri-traversal theory of the organization of mind or for short, a $T_3$-structure. To implement a similar $T_3$-organization in an artificially intelligent agent, it is necessary to have multiple policies, as usually used as a concept in the theory of reinforcement learning. These policies have to form a hierarchy. We define adaptive practopoietic systems in terms of hierarchy of policies and calculate whether the total variety of behavior required by real-life conditions of an adult human can be satisfactorily accounted for by a traditional approach to artificial intelligence based on $T_2$-agents, or whether a $T_3$-agent is needed instead. We conclude that the complexity of real life can be dealt with appropriately only by a $T_3$-agent.

**Introduction: Hierarchy of policies**

Practopoiesis is a theory of how adaptive agents are organized and proposes a number of principles under which such systems operate (Nikolić 2015). One of the key presumptions of practopoiesis is that adaptive mechanisms are organized into a hierarchy: Mechanisms lower on the hierarchy determine the properties of the mechanisms higher on the hierarchy. Interactions among those levels of organization are described by concepts such as monitor-and-act unit and cybernetic knowledge, and by principles such as knowledge extraction, knowledge shielding, downward pressure for adjustment, and equi-level interactions. It has been also proposed that practopoiesis has implications for development of machine learning and artificial intelligence (AI) (Nikolić 2014; Nikolić 2015).

Practopoietic systems can be described from the perspective of machine learning as follows. The entire set of adaptive capabilities of an organism (i.e., monitor-and-act units) at one level of organization, in the terminology of machine learning, can be described as the *policy* ($\pi$) for generating actions. Similarly, cybernetic knowledge (Nikolić 2015) can be understood as an *optimal policy* of machine learning.

Importantly, a practopoietic agent may have different sets of policies, some of them acting on the environment but others acting on the agent itself. These sets form a hierarchy.

Thus, practopoietic hierarchy is an arrangement in which, for policy $x$, there is a policy $y$ whose actions change policy $x$. This makes it a $T_2$-agent, due to the actions executed at two levels of organization. To indicate that actions of policy $y$ change policy $x$, we write:

$\pi_y$ -> $\pi_x$.

In that case, TD-learning (Sutton and Barto 1998) and Q-learning algorithms (Watkins 1989) are considered special policies belonging to $\pi_y$.

Importantly, however, according to practopoietic theory, biological $T_3$-systems have also a third policy (Nikolić 2015)—referred to as tri-traversal theory of human cognition. Thus, a full agent can be described then as follows:

$\pi_G$ -> $\pi_A$ -> $\pi_N$,

whereby, by following the tri-traversal theory, we presume that $\pi_G$ is stored in *genes*, $\pi_A$ in the rules for neural adaptation (responsible also for *anapoiesis*), and $\pi_N$ in the properties of the neural *network*.

To describe the interaction between an agent and its environment we can write:

$\pi_G$ -> $\pi_A$ -> $\pi_N$ -> **U**,

where **U** stands for the surrounding world or Umwelt.

To describe the adaptive capabilities of an entire species, there is one additional policy, $\pi_E$, which determines the genome $\pi_G$. This policy operates according to the rules of *evolution* by natural selection. Thus, the adaptive structure of life on planet Earth can be described as:

$\pi_E \rightarrow \pi_G \rightarrow \pi_A \rightarrow \pi_N$.

That is, the life as a whole has four levels of policies, whereas an individual agent has three.

*Generalizing actions*

In reinforcement learning theory *actions* of an agent are conceptually different from the processes of *learning* by that agent. Practopoiesis generalizes all those forms of actions as adaptive traverses (indicated by arrows). Thus, a system with operational capabilities at $n$ levels is a $T_n$-system and has $n$ traverses, which can be either directed towards the outside of the agent or towards inside of the agent (the latter being often referred to as learning).

The total number of traverses equals the number of organization levels at which policies exist. This is because actions of the policy at the top level of organization, $\pi_N$, affect the environment directly. Hence, the full interaction (with all the arrows) between a biological species as an agent and its environment can be written as:

$\pi_E \rightarrow \pi_G \rightarrow \pi_A \rightarrow \pi_N \rightarrow \mathbf{U}$. (1)

Thus, a species has four traverses, an individual (a biological agent) has three traverses, while an AI agent based on reinforcement learning has only two traverses.

That way reinforcement learning can be considered a special case of practopoietic systems. From the above considerations we can also see that reinforcement learning is not nearly as sophisticated implementation of practopoiesis as is natural intelligence. The reason is more adaptive levels in natural intelligence.

*Generalizing feedback*

In reinforcement learning, learning mechanisms receive feedback. In practopoietic systems, policies at each level of organization receive feedback inputs too. These inputs are conceptually different in reinforcement learning theory: The input for $\pi_x$ is the identity of state $s$; The input for $\pi_y$ is the reward $r$.

Practopoiesis conceptually generalizes those feedback inputs as $i_k$, where $k$ is the level of organization of the system (E, G, A or N). Feedback inputs can be acquired through sensory inputs shared across policies at different levels. For example, a camera can provide the necessary information for $\pi_N$ and $\pi_A$.

The goal of the present study is to demonstrate the advantages of such hierarchical organization of policies as compared to an agent with a single policy.

**Calculating variety of a human agent and its Umwelt**

The problem addressed here is funded in Ashby's (1947) *law of requisite variety*. This law states that for a successful control of a system, the system that controls has to have at least as many states as the system that is being controlled. The question is then how many states can a human brain (or an AI-agent) theoretically assume and is this number sufficiently large to address the variety of the real-life problems that such agents face?

The key presumption behind the present calculations is that the upper limit of the total variety of states that a policy of an agent can produce is related to the total amount of memory of that that policy requires. The available amount of memory represents the maximum entropy that the system can generate and yet that its actions are informed about the environment in which it acts. Therefore, although one may argue that an agent could produce high entropy simply by generating noise, relevant for satisfying the law of requisite variety is only behavior that is informed about the properties of environment. The latter follows from the good regulator theorem (Conant and Ashby 1970), which states that a regulator can be successful in regulating a system only if it is a good model of that system. The memory requirements on variety are the memory requirements for becoming a good model of the world in which the agent operates.

The calculation of total variety gives an upper bound estimate of what an agent can perform, how many different sensory inputs it can distinguish in order to consider making different actions. If one thinks of agent's memory as a set of templates against which the input is matched, then the estimate of total memory is related to an estimate of the number of templates that could be used by that agent.

In other words, we ask the question of how many different patterns (templates) can the brain store in its (synaptic) memory. The number of those patterns indicates the maximum variety of states that the brain, as an Ashby's regulator, can generate for the agent in order to produce meaningful actions on the environment to help with agent's survival. In other words, this number indicates (i) the amount of knowledge that the brain can possibly have on how to respond in a given situation and by doing so, (ii) the variety of responses to sensory inputs that it can produce based on that knowledge.

We are interested only in meaningful informed states i.e., states that reflect some previously acquired knowledge about the surrounding world. Of course the molecules in the brain can have many more states, but if these additional states are not stimulus-dependent they either have to be mutually dependent (correlated) or have to be understood as noise.

We are considering here first agents that do not learn (only later we consider learning). Traditional brain theory and traditional AI both rely on two-traverses (i.e., $T_2$-agents). One of these traverses is for learning, which is the adaptive mechanism located lower on the practopoietic hierarchy. The other traverse is implemented by the mechanisms for processing inputs and executing actions, usually referred to as neural activity in brain sciences and as policies in machine learning. We are interested in the variety of this single top-level traverse. In contrast two the traditional $T_2$-agents, the novel approach with $T_3$-agents presumes the use the two highest-level traverses for processing inputs and executing actions (Nikolić 2015).

The question addressed presently is whether $T_2$-agents can possibly generate sufficient variety of behavior in real-life situations or whether instead only a $T_3$-agent can satisfy those needs. So far, $T_2$-agents have been implemented (either as a brain theory or as an AI) to limited domains of problems, which require much less variety than what an average adult human person may need in real life. The present question is whether these $T_2$-approaches can scale up to the real human-level demands and thus to human-level intelligence.

*Variety generated by an agent*

For an average adult human brain without further learning, we can estimate the total variety based on the total number of synapses and the amount of bits stored at each synapse. According to a recent study, an optimistic estimate is that a single synapse can store 4.6 bits of information (Bartol et al. 2015). Furthermore, if there are about 1000 synapses for each neuron and there are about 100 billion neurons in the brain, we have in total

$10^{11}$ x 1000 = $10^{14}$

or 100 trillion synapses. Given that 4.6 bits of information can be stored per synapse (Bartol et al. 2015), this would set the upper bound of the total theoretical variety that an educated adult human brain can generate without further learning to

4.6 x $10^{14}$ bits.

or

~ 2.4 x $10^{15}$ different states.

This is roughly 500 terabytes of memory, and is within the realms of what can be achieved by today's IT technology.

So, what does that number mean? If synapses were all there is to brain's memory, this is not only the maximum amount of memory stored in the brain at any time, but also the maximum amount of variety that the brain can generate from sensory data. This number gives the maximum number of different responses that the brain can create without further learning. This would also mean that the brain cannot produce a number of internal states larger than about $10^{15}$ that would reflect meaningfully the inputs.

To imagine what this number may indicate, consider the famous patient H.M. who at adult age lost the ability to create new long-term memories after bilateral medial temporal lobectomy (Corkin 1984). H.M. retained all his previous knowledge and was perfectly able to hold a conversation, read text, watch a movie or interact with his environment. He only could not create any new long-term memories. He could not change his knowledge up to the point of the surgery but could perfectly use the previously acquired knowledge. The question is then: how many different stimuli, sentences, events, situations could H.M. distinguish and respond to meaningfully? According to the above calculation and a $T_2$-theory of the brain, this number is $4.6 \times 10^{14}$ bits and indicate the total richness of his mental life that could possibly occur.

In other words, by freezing learning, we are turning a $T_2$-agent into a $T_1$-agent, albeit well trained. This number tells us how many input-output mappings can be maximally preformed.

What the real task behind processing a sensory input is for H.M. (or any other intelligent adaptive agent) is to compare inputs with the entire existing knowledge of all possible patterns that it can detect. In a simple pattern recognition task the agent has to identify the stimulus against its entire database. And we humans can do this very well immedaitelly. For example, we can just *see* a car by checking the shape in the stimulus against all of the other shapes that we have in the memory.

This direct distinguishability of stimuli at the perceptual level for human mind can be tested in experiments with perceptual pop-out (Treisman and Gelade 1980; Treisman 1985). These experiments tell us that we have also limitations. We are not able to distinguish any set of random stimuli, for example: "IOVGJIZGSIOHIO" vs. "IOVGJIKGSIOHIO" cannot be distinguished without a slow serial search for a difference. However, either of the two sequences above can be easily distinguished from: "IOVGJI_SIOHIO". Any every-day visual scene is full of perceptual pop-outs for a human mind.

However, what we humans really excel at in comparison to machines is that we are able to combine this variety of perceptual stimuli with the variety of semantic information. We test everything in parallel, the picture and the its meaning. If the only problem of AI was only finding the difference between two visual stimuli, a simply search algorithm would do that job would by far outperform any human.

Our ability to process semantic information in parallel is estimated by the size of our working memory (or short-term memory). This memory storage is highly limited in capacity (Miller 1956; Cowan 2001), is highly correlated to IQ (e.g., Engle et al. 1999) and is based on semantic information extracted from long-term memory (Miller 1956; Cowan 2001; Olsson and Poom 2005).

We humans can solve many AI-related problems immediately i.e., much like H.M. could, without a need for additional learning. We could just look at a visual scene or just hear a narration and extract much more relevant information than an existing AI-machine can today. We use these simultaneous detection capabilities to make decisions while driving a car, watching a movie, or understanding language, and making purchasing decisions. In all those acts, we compare the current stimulus with all our knowledge acquired until that point in time—and we do it in a blink of an eye.

Thus, the high demands on the variety for an AI-agent come from this parallel template matching against the entire knowledge of the agent. These human capabilities of performing such matching processes fast make us smarter than the machine.

Our superiority is seen most obviously in situation in which the variety of sensory inputs has to be combined with the variety of semantics. This efficient combination of *sensory+semantic* contents makes us much better in understanding visual scenes and natural speech, or simply in playing the game of go.

The present analysis is about the question of whether $10^{15}$ provides sufficient storage for the patterns that the brain needs for such pattern-matching analyses. We compare two different theories of how the brain is adaptively organized ($T_2$ vs. $T_3$ organization) with the estimates of the variety demands posted by the real life of an adult human person.

The present analyses are made under the assumption that all $10^{15}$ combinations are used without any redundancies or other sub-optimalities. Thus, we are estimating the maximal theoretical limits of pattern-matching mechanisms presuming that those have been implemented in the most optimal way possible.

*Variety of real life*

This number $10^{15}$ seems large for producing a lot of intelligent behavior, but the question is: Is it large enough? The other side of the equation is: How much variety does the real life require?

The question of the variety in the real life can be approached by calculating the amount of meaningful variety in sensory inputs to agents. We do not want to estimate the total number of combinations that pixels of an image can assume. We are interested only in the number of combinations that need to be understood by the agent in order to behave successfully in a given world. The question is how many different situations may a human observer need to distinguish, understand and respond to meaningfully? This would be then an estimate of how much variety the human brain should be able to account for.

In the first step of analysis we focus only on the number of different sentences that a human mind may need to be able to comprehend. Our language is generative and a person may expect from the surrounding world any possible message, and should be thus able to decode any of them. To make a rough estimate of the order of magnitude of combinations that can emerge, let us presume that an educated native speaker of English has 15,000 words in a vocabulary (Cervatiuc 2007; Nation and Waring 1997). In addition, let us presume that adverbs, adjectives, verbs and nouns correspond respectively to 5%, 20%, 20% and 55% of the vocabulary. This leads to 750, 3000, 3000, 8250 words in each of the four categories for an average speaker.

From those numbers we can calculate the number of all combinations of sentences of different lengths. For three-word sentences that consists of a noun, followed by a verb and ending with a noun, we obtain roughly:

$8250 \times 3000 \times 8250 \approx 2 \times 10^{11}$

combinations.

This number fits within the variety of the human brain estimated above. But if we add an adjective to each of the nouns noun to make five-word sentences: adjective-noun-verb-adjective-noun, we get a total of

$2 \times 10^{11} \times 3000 \times 3000 \approx 2 \times 10^{18}$

combinations.

This number is already much bigger than the limit that is posed by the total number of synapses in the brain, presuming that synapses are indeed the storage of information and that each synapse can store about 5 bits. If we add an adverb to each verb, the number of combinations grows even further, and so on.

Importantly, it is not clear whether all of these random sentences are meaningful to a human and whether we can consider majority of the combinations as non-meaningful and thus, as not relevant for the variability of a human mind. Indeed, most likely majority of those sentences can be considered meaningless and hence as not being processed by human semantic machinery. To illustrate that point, we list here are a few randomly generated sentences (from http://watchout4snakes.com/wo4snakes/Random/RandomSentence):

"The agony damages the regional spur below a pride."

"Our insult prices the flame."

"Behind the younger textbook quibbles an implied dealer."

But there are also many more six, seven-world long and longer sentences that are meaningful to humans. So, the total number of possible meaningful sentences is not easy to estimate. Before we address this question by another approach—based on the capacity of working memory—let us first point out that $T_2$-theories presume that both sensory and semantic processing are performed at the same level of organization and thus, that it is not just the meaning that the brain (or an AI) needs to account for. It is the also the sensory inputs. All of those functions are covered by the number $4.6 \times 10^{14}$ bits.

This means that the above calculations suggest that a human brain should be unable to distinguish already at the sensory level most of five-word sentences (let alone their meaning). As the discrepancy is not small but is almost four orders of magnitude, this would mean that most of the pairs of random five-word sentences a brain with the variety of 4.6 x $10^{14}$ bits could not be even noticed as two different sentences.

In other words, if we simulate on a computer an artificial neural network with 100 billion neurons, 1000 synapses per neuron and 4.6 bits per synapse, the network would not be large enough to associate a different response for each of the possible five-word sentences but could only do it for three-word sentences.

If the properties of this network correspond to the capacities of our brain, we also could not distinguish most pairs of five-word sentences. Those pairs should sound the same to us if pronounced, or look the same if written on a paper.

But this is clearly not the case. For us, it is easy to distinguish such sentences. How is that possible?

Before addressing this question and discussing the properties of $T_3$-agents, let us first note that the combinatorial problem of the real world vs. the limited variety of a brain, does not stop at language. The problem is the same and becomes possibly even bigger when vision is considered. Vision may require even larger variety than language both at the level of semantics and at the level of sensory inputs. Visual objects have different colors, sizes, shapes, positions, shades, etc.

When trying to understand the variety of processing in vision, we can ask a question of how many meaningful visual scenes our brain is capable of perceiving and distinguishing? To estimated that number, we will turn to the capacity of visual working memory (a.k.a. short-term memory). Working memory is not just a storage of information. It is a place where information is processed and this processing/storage depends primarily on the meaningfulness of the items (e.g., Alvarez and Cavanagh 2004; Olsson and Poom 2005). Working memory stores informaitn by the very means of finding meaning in it (Miller 1956). Hence, the capacity of working memory can be used as an indicator of how much meaning can visual system extract from a visual scene.

Experiments indicate that visual working memory can store about four objects (Luck and Vogel 1997) and only if we are very familiar with them (Cowan 2001; Alvarez and Cavanagh 2004; Nikolić and Singer 2007) and only if a category exists for each object (Olsson and Poom 2005). Thus, if we conservatively assume that an adult human is able to distinguish 10,000 different categories of objects, working memory for four objects would require a total variety of

$$(10^4)^4 = 10^{16}$$

combinations. This would mean that already the combinations needed for visual working memory cannot be accounted for by the memory of $10^{15}$ states. WM capacity reflects human capacity to understand a visual scene and is tightly related to the attentional capacity (Alvarez and Cavanagh 2004; Awh and Jonides 2001; Mayer et al. 2007). The present result would mean that the variety provided by our synaptic memory is not sufficient to enable us to understand a visual scene of four objects.

One possibility is that a $T_2$-brain has more capacity to generate variety than the currently estimated. Another possibility would be that the capacity of four is an overestimate and involves some type of chunking (as shown for task that show capacity larger than four; Cowan 2001) and that the "true" capacity of visual working memory is perhaps just three objects. The latter hypothesis would lead to

$$(10^4)^3 = 10^{12}$$

combinations, and would fit well within the supposed $10^{15}$ combinations of a $T_2$-brain. Therefore, similarly to what we have concluded for the semantics of verbal materials, the semantic properties of visual working memory may fit—with some stretching(!)—to the apparent limits of the brain.

However, even if both of the above hypotheses were correct and the brain had in the same time more storage than assumed (e.g., more synapses) and the working-memory capacity of three objects, not four, still another source of a combinatorial problem would remain. The above calculation accounts only for the semantic memory i.e., object identities, and does not take into account the variety of sensory inputs with which these objects come. The fact is that there is not a single shape, size, color or shading for most of the objects that we can recognize and categorize. Normally, visual objects come in a huge variety of visual appearances and this variety needs also to be taken care of by the brain.

If we conservatively assume that we can perceptually easily detect each object in just 10,000 different forms, this leads to:

$$(10^{12})^4 = 10^{48}$$

combinations for three-object working memory (attention capacity), and to

$$(10^{16})^4 = 10^{64}$$

combinations for four objects. These numbers exceed readily the estimated capacity of the brain.

In fact, the number of visual combinations in which visual objects can come and can be perceptually distinguished by our visual system without any significant effort may be even larger. If we just assume that we can perceive an object—such as e.g., a car—in 10 different shapes, in 10 sizes of retinal projections and in 10 orientations, with 10 different colors, and 10 patterns of shading, we already have $10^5$ combinations for that object. And these numbers are likely to be much higher in reality. A similar problem holds for auditory inputs and recognition of speech.

These real-life variety numbers seem too high to be accounted for by stretching the estimates of the number of synapses or their individual memory capacity. Rather, it seems that there is a fundamental discrepancy between what a $T_2$-brain of reasonable size can offer (be it biological or not) and what the real-life demands pose on human-level intellectual capabilities.

In conclusion, it seems that the $T_2$-theory of the brain, which bases mental operations on a single policy, may account for the total variety of semantics, but the problem is with the additional variety of perceptual inputs. It seems that the combinatorial possibilities of perceptual inputs in real life create the real problem as they need to combine with semantics and the resulting variety exceed by far what a maximally optimized brain with 100 billion neurons and 1000 synapses per neuron could possibly deal with.

*Variety of $T_3$-agents*

The above problems have been encountered when a single policy was considered. Here, we will discuss how multiple policies can provide a relief for that problem (called *variety relief* in practopoietic theory; Nikolić 2015). To understand the solution offered by the variety relief in $T_3$-agents, it is useful to first consider the boost in variety that can be achieved by the process of learning in a $T_2$-agent. If learning is not frozen and thus, we presume a full healthy brain (not H.M.'s brain), we can repurpose the resources and replace one type of knowledge that is no longer needed, with new knowledge that may be more valid in a new situation. That way, when learning is allowed, a much higher total variety can be produced.

For example, if memory storage for some text-storage device is limited to just one million characters, only one or a few books can be stored in this memory. However, if the device can "relearn" by deleting old and loading new books, the device can store all possible books that do not exceed 1 million characters. In fact, the total possible variety of that memory storage for is $10^{156}$ of different combinations of 26 letters in English alphabet (for comparison, the number of atoms in the visible universe is $\sim 10^{80}$).

With a limited brain size or neural-network size, changes to the network's knowledge are thus the key process for boosting variety.

But what if not only the slow learning of facts and skill boosts variety, but in addition another mechanism operates and makes the brain change its knowledge at another level and at high speed. If the brain would have some quick way of reorganizing its anatomy and changing its memories, say in less than a second, it could produce a much higher variety than $10^{15}$. It may have in fact enough variety to account for the richness of the sensory inputs.

The hierarchy of policies in a $T_3$-agent described above in (1) can offer exactly this learning-based boost in variety. As policy $\pi_A$ can change policy $\pi_N$, and the total variety of the agent increases.

*How much can the variety increase theoretically?*

We have seen that maximum possible variety of a 1-million character storage is $10^{156}$, and this puts the upper bound as it presumes the "learning" mechanism (i.e., the loading mechanism), that is itself unlimited in knowledge creation capabilities. However, in most cases this is not realistic. The learning mechanism has its own limitations.

In general, when the variety of the learning mechanism is considered, the combined variety across two levels of organization can be computed as a product of the two varieties: If $\pi_A$ has $N_A$ possible states, and $\pi_N$ has $N_N$ states, the maximum total theoretical number of states that could be produced by the combined agent is:

$N_A \times N_N$.

For example, in the 1-million character memory from the example above we may presume a book-loading "learning" mechanism that has only 10 different states; this loading mechanism cannot load more than 10 different books. As a consequence, the total possible variety of the entire system (*memory* + *loader*) is 10 million different states. In general, depending on the limitations of the learning mechanisms, there will be normally a stark reduction in the number of combinations in comparison to what would be achieved by an unlimited learning mechanism (in the above example $10^7$ down from $10^{156}$).

In adaptive systems, the limitations on learning come from the limited sources of knowledge. If the knowledge would be already prepared in a ready-to-use form and stored elsewhere, it could be simply loaded (like from a larger hard-disk to the smaller RAM memory of a computer). This would make the problem trivial. Unfortunately, adaptive systems do not have such an auxiliary depository of knowledge of how to interact with the world. Rather, biological systems have to extract that knowledge from the environment, which is why they are *adaptive* on the first place.

As mentioned, the process of extracting knowledge from the environment is referred to as traverse in Nikolić (2015). For example, application of reinforcement learning is a traverse; knowledge on how to learn stored at a lower level is applied through interaction with the environment in order to create new knowledge (new policy) at a higher level. Hierarchy of policies in (1) generalizes that relation.

*How many states can a frozen $T_3$-brain theoretically produce?*

Let us presume that the brain is a $T_3$-agent and that when frozen (i.e., without learning), it becomes a $T_2$-agent. Let us also presume that the brain uses much of its variety for the lower level of the two remaining, i.e., for storing $\pi_A$. This is where the abstract knowledge is stored such as concepts. Hence, this level of organization can be referred to as *ideatheca* (meaning *storage of concepts*).

Let us conservatively assume that ideatheca (i.e., $\pi_A$) has just $10^{12}$ states, which is what we estimated above as the lower bound of semantic capacity enabling three-item working-memory. Next, let us presume that $\pi_N$ has even less variety and set it to the value $10^{10}$. This presumes that only a small portion of the entire brain's resources is under the influence of ideatheca and can be changed quickly in less than a second. In particular, the choice of this number presumes that only 1/1000-th of the total memory machinery of the brain is being changed in such a rapid way.

Under these assumptions, the total number of states that a combined $\pi_A \rightarrow \pi_N$ could produce without any additional learning is:

$10^{12} \times 10^{10} = 10^{22}$.

This number is much larger than $10^{15}$ and much more suitable for coping with the estimated real-life requirements on variety. This number indicates that if H.M. was a $T_3$-agent before the surgery and became limited to a $T_2$-agency after the surgery (loosing his third traverse), this patient may have had the possible richness of mental life that could deal with $10^{22}$ combinations. Irrespective of whether the estimates of his semantic memory of concepts is about $10^{12}$ or $10^{16}$, there is still a lot of room left for additional combinations of sensory inputs that indicate those concepts in the surrounding world and that H.M. could efficiently process.

The number $10^{22}$ would also correspond to a neural network that has 100 billion neurons and 1000 synapses per neuron, but also has an additional set of mechanisms that change the properties of the network with a rapid rate and on the basis of the incoming sensory inputs. To achieve variety of $10^{22}$, it would be sufficient to enable changing one bit of information per neuron (there are about 100 billion neurons in human brain). For example, a neuron could be switched on or off by its adaptation mechanisms.

For this to work, a pre-requirement is that the slow learning mechanisms noted as $\pi_G$ in (1) provide the knowledge to $\pi_A$ on how to adjust $\pi_N$. In other words, by slow learning mechanisms and throughout many years of the development of the nervous system the network must first learn how to make these quick adjustments to its $\pi_N$. That is the network has to acquire the $10^{12}$ amount of $\pi_A$ knowledge through its development time.

In that case the agent can be considered as "understanding" the sensory inputs. Understanding would mean that the operation of $\pi_A$ give the stimuli best possible interpretation given all of the knowledge that the agent has acquired through lifetime (for details see the section on abductive reasoning in tri-traversal agents in Nikolić 2015).

The alternative to extending the hierarchy would be to cope with the variety requirements by simply increasing the total size of the given policy i.e., by increasing the network size. In that case, variety grows linearly with the number of components; to double the variety of patterns stored in the brain, the size of the brain needs to be doubled. To increase variety to $10^{22}$ states, from $10^{15}$ states in a 1.5-kilogram brain, we would need an increase to $1.5 \times 10^7$ kilograms of biological mass. This is more than the cumulative size of all the brains of all the people currently living on planet Earth.

**Conclusions**

A $T_2$-AI, which means an AI based on single memory storage and on a single set of learning mechanisms, cannot possibly reach the intelligence of human. This conclusion is made on the basis of Ashby's (1947) requisite variety theorem and an estimate of the total theoretical variety of control that a brain can create given the number of neurons and synapses. It turns out that the variety the brain could possibly create if it was a $T_2$-agent would be to deal successfully with the demands of a real-world environment.

However, we also show that if the organization of the brain formed a $T_3$-agent, sufficient variety would be generated for dealing with real life. Accordingly, an AI that would mimic human intelligence would have to be organized as a $T_3$-agent.

The variety of a $T_2$-agent would be sufficient to implement all of the semantic knowledge of an adult human person, but would not suffice for the requirements of the sensory processing of those semantic categories. The objects and situations that need to be detected from the sensory data, be it recognition of visual scenes or understanding speech, require too much variability to be dealt with a $T_2$-brain, even if the coding and processing is maximally optimized in this brain.

The important implication of the present analysis is that no novel optimization or invention of a new algorithm, or discovery of a new architecture for neural circuits can possibly bring a $T_2$-agent (i.e., a traditional single-policy + learning-mechanism agent) with reasonable size of resources to a human-level intelligence. The present calculations already presume that all the operations and coding schemes in the organization of the agent have been optimized to the theoretical maximum. Thus, no new creative invention in machine learning is possible that could bring to the intelligence level of humans the classical approaches to AI. In other words, to built artificial general intelligence we need to seek beyond deep learning networks, Markov chains or Bayesian networks. Otherwise, we would need to scale up the resources to prohibitively large sizes.

The only way to create an artificial system that is human-level intelligent with reasonable resources is to implement a hierarchy of policies, which then makes possible the decisions about driving, walking, moving etc. to rely on the full variety of the sensory data. A $T_3$-agent with realistic computational resources can perform such a task and, once it has acquired knowledge of an average adult person, it could generate variety of $10^{22}$ states. This number is sufficient to deal with all the semantic knowledge and still plenty of sensory information can be processed. And, if needed, there would be enough room for increasing that number within the realm of the current IT technology.

A change from $T_2$ to $T_3$-organization comes with some costs (Nikolić 2015). One cost is that the entire agent operates always slower with more than with fewer traverses. This is because the additional adaptive processes require time to complete. In human mental operations, this slowdown ranges from 100s of milliseconds or seconds (for more details see Nikolić 2015).

In Nikolić (2015) it has been proposed that the physiological mechanism underlying anapoiesis, i.e., the application of knowledge in ideatheca to change network properties, are implemented through neural adaptation. Furthermore, these mechanisms are proposed to rely on sensory inputs and hence, largely on the variety stored in synapses that process those sensory inputs.

In conclusion, an AI that matches human intellectual capabilities is possible only in tri-traversal systems.

**Acknowledgments**

The author would like to thank Hrvoje Nikolić, Raul C. Muresan and Shan Yu for valuable comments on previous versions of the manuscript.